\pdfoutput=1

\documentclass[aps,prd,preprint,groupedaddress]{revtex4}

\usepackage[english]{babel}
\usepackage{graphicx}

\usepackage{amsmath,amsfonts,amssymb}
\usepackage{calrsfs}     
\usepackage{slashed}     

\newcommand\comments[1]{}   

\begin{document}

\title{One light composite Higgs boson \\ facing electroweak precision tests}

\author{Marc Gillioz}
\email{gillioz@physik.uzh.ch}
\affiliation{Institut de Th\'eorie des Ph\'enom\`enes Physiques, EPFL, \\ CH-1015 Lausanne, Switzerland}
\affiliation{Institut f\"ur Theoretische Physik, Universit\"at Z\"urich, \\
Winterthurerstrasse 190, CH-8057 Z\"urich, Switzerland}

\date{\today}

\begin{abstract}
	We study analytically and numerically the bounds imposed by the electroweak precision tests on a minimal composite Higgs model. The model is based on spontaneous $SO(5) \rightarrow SO(4)$ breaking, so that an approximate custodial symmetry is preserved. The Higgs arises as a pseudo-Goldstone boson at a scale below the electroweak symmetry breaking scale. We show that one can satisfy the electroweak precision constraints without much fine-tuning. This is the case if the left-handed top quark is fully composite, which gives a mass spectrum within the reach of the LHC. However a composite top quark is strongly disfavoured by flavour physics. The alternative is to have a singlet top partner at a scale much lighter than the rest of the composite fermions. In this case the top partner would be light enough to be produced significantly at the LHC.
\end{abstract}

\maketitle


\section{Introduction}

The aim of this paper is to study quantitatively the constraints of the electroweak precision tests (EWPT) on a class of composite Higgs models \cite{Kaplan:1983fs} in which the Higgs arises as a pseudo-Goldstone boson of some strongly interacting sector. In these models, a global symmetry $G$, in which the Standard Model (SM) symmetry is embedded, is spontaneously broken to a subgroup $G'$, yielding a definite number of Goldstone bosons, among which the $SU(2)_L$ doublet is identified as the usual Higgs. The physical Higgs boson acquires a mass through radiative corrections from an additional explicit breaking of $G$. The scale of the spontaneous breaking $G \rightarrow G'$ is denoted by $f$, whereas the electroweak symmetry breaking (EWSB) is generated at a lower scale, $v=247.5$~GeV. The hierarchy between those two scales can be expressed in term of a parameter $\epsilon = v / f \lesssim 1$.

The first composite Higgs models derived more than 20 years ago \cite{Kaplan:1983fs} were problematic because of the presence of flavour changing neutral currents (FCNC), and the lack of calculability. It has been recently shown that models with a warped fifth dimension lead to a similar effective 4D theory, but with a different realisation of the fermions, avoiding FCNC. Moreover, these models become calculable in the limit of a large number of colors. In this paper we will consider a 4D low-energy effective description of a 5D model approximately preserving the custodial $SO(4)$ symmetry \cite{Contino:2006qr,Carena:2006bn}. In particular we consider a two-site model \cite{Contino:2006nn} where only the first resonance of the full tower of Kaluza-Klein modes appears. The simplest possible extension of $SO(4)$ yielding 4 Goldstones when broken is $SO(5)$, so we take the latter to be the spontaneously broken global symmetry $G$. The importance of custodial symmetry comes from the EWPT and in particular from the necessity to protect the Peskin-Takeuchi $T$ parameter \cite{Peskin:1990zt} from large contributions at tree level.

It has been stressed by Barbieri et al.~\cite{Barbieri:2007bh} that an important correction to the Peskin-Takeuchi $S$ and $T$ is generated in this class of composite models by a shift $\sqrt{1 - \epsilon^2}$ in the coupling of the Higgs to any other field. This is equivalent to having an effective mass for the Higgs:
\begin{equation}
	m_{H, \text{eff}} = m_H \left( \frac{\Lambda}{m_H} \right)^{\epsilon^2},
	\label{equ:mHeff}
\end{equation}
where $\Lambda$ is the cutoff of the theory. This infrared correction to $S$ and $T$ makes the theoretical predictions not compatible with the experimental data for a reasonable choice of $\epsilon$ (see figure~\ref{fig:e1e3}). Nevertheless, the agreement can be re-obtained if an extra positive contribution to $T$ from the new states is present. But as soon as the contribution to $T$ becomes important, one should keep control of the corresponding correction to the decay $Z \rightarrow b \bar b$, which is also strongly constrained by the electroweak precision measurements. In the following, we propose to make a complete and quantitative computation of the EWPT in order to evaluate the bounds on the parameters of the model.


\section{A minimal model}

The minimal model of a composite Higgs preserving custodial symmetry is based on $SO(5) \rightarrow SO(4)$ breaking. An additional $U(1)_X$ local symmetry is needed to give the correct hypercharge to the fermions. Since $SO(4)$ is isomorphic to $SU(2)_L \times SU(2)_R$, one can choose the SM gauge group factors $SU(2)$ and $U(1)_Y$ to be respectively $SU(2)_L$ and an admixture of the third component of $SU(2)_R$ and the X-hypercharge.

The Goldstone bosons come from a scalar field $\Sigma$ transforming like a $\textbf{5}$ of $SO(5)$, uncharged under $U(1)_X$. This scalar field is taken to be dimensionless and subject to the normalised constraint:
\begin{equation}
	\Sigma^2 = 1.
\end{equation}
One can expand $\Sigma$ around a $SO(4)$-preserving vacuum state $\Sigma_0 = (0,0,0,0,1)$ as
\begin{equation}
	\Sigma = \Sigma_0 ~ e^{-i T^{\hat a} h^{\hat a} \sqrt 2 / f},
\end{equation}
where $T^{\hat a}$ are the broken $SO(5)/SO(4)$ generators given by:
\begin{equation}
	T^{\hat a}_{ij} = - \frac{i}{\sqrt 2} \left( \delta^{\hat a}_i \delta^5_j
		- \delta^5_i \delta^{\hat a}_j \right),
	~~~~~~ \hat a = 1,2,3,4.
\end{equation}
We denote by $\vec \Sigma$ the first four components of $\Sigma$, which is an $SO(4)$ symmetric vector, and by $\Sigma_5$ its fifth component, so that we have:
\begin{equation}
	\Sigma = \left( \vec\Sigma, \Sigma_5 \right) = \left(  \frac{h^{\hat a}}{h} \sin\frac{h}{f}, \cos\frac{h}{f} \right),
	~~~~~~ h = \sqrt{(h^{\hat a})^2}.
\end{equation}

In the most general model many fermionic and vectorial resonances may appear, but since we focus on low-energy phenomenology, we take into account the first fermionic resonance $\chi$ only. We also restrict the elementary SM fermions to the third generation, namely $q_L = (t_L, b_L)$, $t_R$, $b_R$, since the contribution to the EWPT from the first two generations is negligible. $\chi$ is a $\textbf{5}$ of $SO(5)$, broken down to a bidoublet of $SU(2)_L \times SU(2)_R$ and a singlet: $\textbf{5} = (\textbf{2}, \textbf{2}) \oplus (\textbf{1}, \textbf{1})$. It is charged under $U(1)_X$ with hypercharge $X=\frac{4}{3}$. Under the SM gauge group, $\chi$ decomposes as:
\begin{equation}
	\chi =
	\left(\begin{array}{c}
		Q \\ X \\ \tilde T
	\end{array}\right),
	~~~~~~~~
	Q = \left(\begin{array}{c}
		T \\ B
	\end{array}\right),
	~~~~
	X =
	\left(\begin{array}{c}
		X_{5/3} \\ X_{2/3}
	\end{array}\right),
\end{equation}
where $\tilde T$ has the same quantum numbers as the right-handed top quark $t_R$, the $Q$ doublet has the same quantum numbers as $q_L$, and $X$ is made of two fermions of electric charge $\frac{2}{3}$ and $\frac{5}{3}$. The mass mixing between the elementary and composite sectors is achieved through the simplest linear mixing:
\begin{eqnarray}
	\mathcal{L} &=& i \bar q_L \slashed D q_L + i \bar t_R \slashed D t_R
		+ i \bar b_R \slashed D b_R \nonumber \\
	&& + \bar\chi \left( i \slashed D - M_0 \right) \chi
		+ \frac{1}{2} f^2 \left( D_\mu \Sigma \right) \left( D^\mu \Sigma \right)
		- y^* f~\bar\chi_i \Sigma_i~\Sigma_j \chi_j \label{equ:Lagr} \\
	&& + \left( \Delta_L \bar q_L Q_R + \Delta_R \bar T_L t_R + \text{h.c.} \right). \nonumber
\end{eqnarray}
Notice that we have omitted the SM gauge bosons and the composite vector resonances which may appear. In the following we will also take into account a vector field $\rho$ in the adjoint representation, a $\textbf{10}$ of $SO(5)$.

The theory defined by this Lagrangian is not renormalisable in 4D, but as we mentioned before it is fully satisfactory in the 5D picture. The diagonalisation is achieved by mixing the elementary fermions with their composite partners. In particular, for $\Sigma=\Sigma_0$, the Lagrangian becomes diagonal after the rotations:
\begin{eqnarray}
	\left(\begin{array}{r}
		q_L \\ Q_L
	\end{array}\right)
	\rightarrow
	\left(\begin{array}{rr}
		\cos \phi_L & -\sin \phi_L \\
		\sin \phi_L & \cos \phi_L
	\end{array}\right)
	\left(\begin{array}{r}
		q_L \\ Q_L
	\end{array}\right)
	&~~&
	\tan \phi_L = \frac{\Delta_L}{M_0},
	\label{equ:phiL} \\
	\left(\begin{array}{r}
		t_R \\ \tilde T_R
	\end{array}\right)
	\rightarrow
	\left(\begin{array}{rr}
		\cos \phi_R & -\sin \phi_R \\
		\sin \phi_R & \cos \phi_R
	\end{array}\right)
	\left(\begin{array}{r}
		t_R \\ \tilde T_R
	\end{array}\right)
	&~~&
	\tan \phi_R = \frac{\Delta_R}{\tilde M_0},
	\label{equ:phiR}
\end{eqnarray}
where we have defined:
\begin{equation}
	\tilde M_0 = M_0 + y^* f.
	\label{equ:Mtilde}
\end{equation}
Here $\phi_L$ and $\phi_R$ parametrise the degree of compositeness of the left- and right-handed top quark respectively. In this regime, the mass of the two heavy doublets $Q$ and $X$ is close to $M_0$, whereas the right-handed top partner $\tilde T$ appears at a different scale $\tilde M_0$. Both $Q$ and $\tilde T$ have their mass increased by level repulsion since they are mixing with the SM fields $q_L$ and $t_R$:
\begin{eqnarray}
	m_Q & = & \frac{1}{\cos \phi_L} M_0 \label{equ:mQ}, \\
	m_X & = & M_0, \\
	m_{\tilde T} & = & \frac{1}{\cos \phi_R} \tilde M_0. \label{equ:mT}
\end{eqnarray}

Since the linear coupling $\Delta_L$ and $\Delta_R$ break the $SO(5)$ symmetry explicitly, a potential that removes the Goldstone degeneracy is generated \footnote{Other effects may also be needed to obtain a realistic potential for the scalar.} and $\Sigma$ gets a vacuum expectation value (VEV) different from $\Sigma_0$:
\begin{equation}
	\langle \vec \Sigma^2 \rangle = \epsilon^2 = \frac{v^2}{f^2},
	~~~~~~
	\langle \Sigma_5 \rangle = \sqrt{1 - \epsilon^2}.
\end{equation}
In this case, the diagonalisation becomes more complicated than equations (\ref{equ:phiL}-\ref{equ:phiR}) since it involves a $4\times 4$ matrix. The masses of the composite top partners $T$, $X_{2/3}$ and $\tilde T$ get corrected by an amount proportional to $\epsilon^2$, while the elementary top quark obtains a mass, given at the leading order in $\epsilon$ by:
\begin{equation}
	m_t = \frac{\epsilon}{\sqrt 2} ~ \sin \phi_L ~ \sin \phi_R ~ y^* f. 
	\label{equ:topmass}
\end{equation}

The custodial $SO(4)$ symmetry is broken only by the $\Delta_L$ operator of equation (\ref{equ:Lagr}), and therefore we expect the contribution from new physics to Peskin-Takeuchi $T$ parameter and to the $Z \rightarrow b \bar b$ to be of order $\epsilon^2$ times the SM values.

Finally, requesting the mass of the top quark to be its measured value $m_t = 172.5 \pm 2.7$~GeV \cite{Yao:2006px}, the model can be fully parametrised in terms of four dimensionless parameters: $\epsilon$, $\phi_L$, $\phi_R$ and the ratio $\tilde M_0 / M_0$.


\section{Electroweak precision tests}

\begin{figure*}[t]
	\centering
	\parbox{.46733\linewidth}{\includegraphics[width=\linewidth]{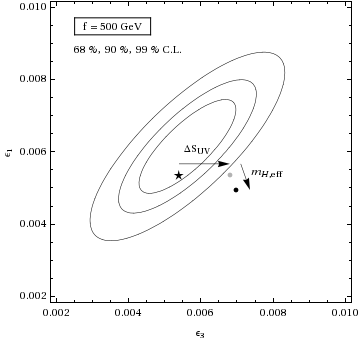}} 
	\parbox{.48\linewidth}{\includegraphics[width=\linewidth]{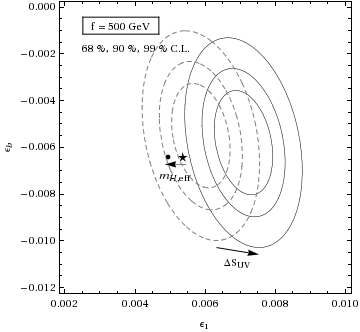}}
	\caption{Left: allowed region in the plane $(\epsilon_3,\epsilon_1)$ fixing $\epsilon_2$ and $\epsilon_b$ to their SM values; the star corresponds to the Standard Model prediction, the black dot to our composite model. Right: same in the plane  $(\epsilon_1,\epsilon_b)$ with $\epsilon_2$ and $\epsilon_3$ fixed to the SM values (dashed ellipses) and with the UV correction to $\epsilon_3$ turned on (solid ellipses).  Here we have used $m_H=120$~GeV, $m_\rho=2.5$~TeV and $f=500$~GeV ($\epsilon^2 \approx 0.25$).}
	\label{fig:e1e3}
\end{figure*}
As we mentioned above, important constraints on the parameters of this model come from the EWPT, and in particular from the oblique corrections associated with $S$ and $T$, which are strongly bounded by LEP experiments. Since the top Yukawa $\lambda_t$ is much larger than the other Yukawa couplings and also larger than the gauge couplings $g$ and $g'$, the correction to $T$ from new physics will be dominated by a loop of third-generation fermions, namely the elementary $t$ and the composite $\chi$. On the other hand, the top partners will not contribute significantly to $S$. The main correction to it comes from the composite vector resonance $\rho$ at tree-level. It can be evaluated quantitatively using the 5D picture of this model as a function of the $\rho$ mass. In our case, considering only the first resonance instead of the whole Kaluza-Klein tower, we have \cite{Giudice:2007fh,Contino:2008}:
\begin{equation}
	\Delta\hat S_\text{UV} = \frac{\alpha}{4 s_W^2} \Delta S_{UV} \cong \frac{m_W^2}{m_\rho^2} \left( 1 + \frac{m_\rho^2}{m_a^2} \right),
\end{equation}
where $m_a$ is the mass of the $SO(5)/SO(4)$ vectors. Using the 5D picture, we can compute it to be $m_a \cong 5/3~m_\rho$ \cite{Agashe:2004rs}. The mass of the $\rho$ field is directly proportional to the scale of new physics: $m_\rho = g_\rho f$. The coupling $g_\rho$ is roughly imposed by the large number of colors in the composite sector: for $f = 500$~GeV, we fix the mass of the vector resonance to $m_\rho = 2.5$~TeV \cite{Agashe:2004rs}. Hence $g_\rho = 5$ and we have:
\begin{equation}
	\Delta \hat S_\text{UV} \cong \left( 5.7 \cdot 10^{-3} \right) \epsilon^2. \label{equ:Scorrection}
\end{equation}
For a significant $\epsilon^2$, this contribution to $S$ is large but may still be compatible with the experimental data provided that there is also a large positive contribution from the new physics to $T$. This last point may be problematic since $T$ is related to the correction of the $Z \rightarrow b_L \bar b_L$ vertex, which is also strongly constrained by measurements. This non-oblique correction is denoted by $\tau$ and is defined as the modification of the coupling of the $Z$ boson to left-handed bottom quarks:
\begin{equation}
	\mathcal{L}_\text{eff} \supset i \frac{g}{2 c_W}
		\left( 1 - \frac{2}{3} s_W^2 + \tau \right)
		Z_\mu \bar b_L \gamma^\mu b_L.
\end{equation}
Like for $T$, the main contributions to $\tau$ come from the top partners, due to the large top Yukawa coupling.

The oblique corrections to the electroweak precision observables coming from the SM and from the new physics can be expressed in terms of three parameters slightly different from Peskin-Takeuchi ones: $\epsilon_1$, $\epsilon_2$, $\epsilon_3$ \cite{Altarelli:1990zd}. In addition one can define $\epsilon_b$ in order to describe the bottom quark sector as well \cite{Altarelli:1993sz}. The advantage of those new parameters is that they are directly related to experiment and do not depend on the model, unlike the Peskin-Takeuchi ones, which are defined at a fixed point in the Standard Model. $\epsilon_1$ and $\epsilon_3$ are closely related to $T$ and $S$ respectively, while $\epsilon_b$ depends directly on the correction to the $Z \rightarrow b_L \bar b_L$ vertex, $\tau$:
\begin{eqnarray}
	\epsilon_1 & = & \left( +5.60 - 0.86 \log \frac{m_H}{m_Z} \right) \cdot 10^{-3}
		+ \Delta \epsilon_1^\text{IR} + \Delta\hat T,
		\label{equ:e1th} \\
	\epsilon_2 & = & \left( -7.09 + 0.16 \log \frac{m_H}{m_Z} \right) \cdot 10^{-3},
		\label{equ:e2th} \\
	\epsilon_3 & = & \left( +5.25 + 0.54 \log \frac{m_H}{m_Z} \right) \cdot 10^{-3}
		+ \Delta \epsilon_3^\text{IR} + \Delta\hat S,
		\label{equ:e3th} \\
	\epsilon_b & = & -6.43 \cdot 10^{-3} + \Delta \tau.
		\label{equ:ebth}
\end{eqnarray}
The numerical values are those obtained by computing the SM corrections \cite{Agashe:2005dk}. $\Delta\hat T$, $\Delta\hat S$ and $\Delta \tau$ are the contributions from the new physics to $T$, $S$ and $\tau$ coming from the new heavy states. In that sense they are ultraviolet corrections. $\Delta \epsilon_1^\text{IR}$ and $\Delta \epsilon_3^\text{IR}$ are the infrared corrections coming from the effective mass of the Higgs as in equ.~(\ref{equ:mHeff}). They are given by \cite{Peskin:1990zt}:
\begin{eqnarray}
	\Delta \epsilon_1^\text{IR} & = & -\frac{3}{16 \pi c_W^2} \alpha ~ \epsilon^2 \log \frac{m­_\rho^2}{m_H^2}
		\label{equ:delateps1}, \\
	\Delta \epsilon_3^\text{IR} & = & \frac{1}{12 \pi} \frac{\alpha}{4 s_W^2} \epsilon^2
		\log \frac{m­_\rho^2}{m_H^2}. \label{equ:delateps3}
\end{eqnarray}

Experimentally, the $\epsilon$'s are given by the LEP experiment \cite{Agashe:2005dk, Barbieri:2004qk}:
\begin{equation}
	\begin{array}{rcl}
		\epsilon_1 & = & \left( +4.9 \pm 1.1 \right) \cdot 10^{-3} \\
		\epsilon_2 & = & \left( -9.1 \pm 1.2 \right) \cdot 10^{-3} \\
		\epsilon_3 & = & \left( +4.8 \pm 1.0 \right) \cdot 10^{-3} \\
		\epsilon_b & = & \left( -5.2 \pm 1.5 \right) \cdot 10^{-3}
	\end{array}
	~~~~~~
	\rho =
		\left(\begin{array}{cccc}
			1 & 0.59 & 0.83 & -0.28 \\
			0.59 & 1 & 0.45 & -0.13 \\
			0.83 & 0.45 & 1 & -0.16 \\
			-0.28 & -0.13 & -0.16 & 1
		\end{array}\right),
	\label{equ:epsilons}
\end{equation}
where $\rho$ is the correlation matrix of the four $\epsilon$'s. Fig.~\ref{fig:e1e3} shows the relation between the experimental constraints and the predictions of our model, compared to the SM.

The new physics corrections $\Delta\hat T$ and $\Delta \tau$ can be computed explicitly at one loop as a function of the four free parameters of the model. Since the top has three composite partners which have the same electric charge, the complete computation requires the diagonalisation of a $4 \times 4$ mass matrix. One way to proceed analytically is to expand $\Delta\hat T$ and $\Delta \tau$ as a power series in $\epsilon$. At leading order we recover the SM top loop corrections, since for $\epsilon$ very small all the new states are really heavy and decouple:
\begin{eqnarray}
	\hat T_\text{SM}^\text{top} = \frac{3 m_t^2}{16 \pi^2 v^2} \cong 9.2 \cdot 10^{-3}, \\
	\tau_\text{SM}^\text{top} = -\frac{m_t^2}{8 \pi^2 v^2} \cong -6.2 \cdot 10^{-3}.
\end{eqnarray}
At next-to-leading order in $\epsilon$, the new physics contributions are in a complicated form, but one can estimate their importance by taking appropriate limits. For $\tilde M_0 / M_0 \rightarrow 0$, the doublets are much heavier than the singlet and hence only $\tilde T$ contributes ($m_Q, m_X \gg m_{\tilde T}$) and we have:
\begin{eqnarray}
	\Delta\hat T_{\tilde T} & = & \hat T_\text{SM}^\text{top} \left[ 2 \frac{m_t^2}{m_{\tilde T}^2}
		\frac{1}{\tan^2 \phi_R} \left( \log \frac{m_{\tilde T}^2}{m_t^2} - 1
		+ \frac{1}{2 \tan^2 \phi_R} \right)
		\right],
	\label{equ:TT} \\
	\Delta \tau_{\tilde T} & = & \tau_\text{SM}^\text{top} \left[ 2 \frac{m_t^2}{m_{\tilde T}^2}
		\frac{1}{\tan^2 \phi_R} \left( \log \frac{m_{\tilde T}^2}{m_t^2} - 1
		+ \frac{1}{2 \tan^2 \phi_R} \right) \right].
	\label{equ:tauT}	
\end{eqnarray}
In the limit $\tilde M_0 / M_0 \rightarrow \infty$, only the two doublets $Q$ and $X$ contribute. We can look at two particular cases: first, if $\sin \phi_L$ is large, $Q$ is much heavier than $X$ ($m_{\tilde T}, m_Q \gg m_X$) and we have:
\begin{eqnarray}
	\Delta\hat T_{X} & = & \hat T_\text{SM}^\text{top} \left[ -4 \frac{m_t^2}{m_X^2}
		\left( \log \frac{m_X^2}{m_t^2} - \frac{11}{6} \right) \right],
	\label{equ:TX} \\
	\Delta \tau_X & = & \tau_\text{SM}^\text{top} \left[ -\frac{1}{2} \frac{m_t^2}{m_X^2}
		\log \frac{m_X^2}{m_t^2} \right].
	\label{equ:tauX}
\end{eqnarray}
Conversely, if $\sin \phi_L$ is close to zero, the mass splitting between $X$ and $Q$ is small ($m_{\tilde T} \gg m_Q, m_X$ and $m_Q \sim m_X$), and we obtain approximately twice the contributions of the doublet $X$ alone:
\begin{eqnarray}
	\Delta\hat T_{Q, X} & = & \hat T_\text{SM}^\text{top} \left[ -8 \frac{m_t^2}{m_X^2}
		\left( \log \frac{m_X^2}{m_t^2} - \frac{143}{60} \right) \right],
	\label{equ:TQX} \\
	\Delta \tau_{Q, X} & = & \tau_\text{SM}^\text{top} \left[ - \frac{m_t^2}{m_X^2}
		\left( \log \frac{m_X^2}{m_t^2} - \frac{1}{2} \right) \right].
	\label{equ:tauQX}
\end{eqnarray}
Equ.~(\ref{equ:TT}), (\ref{equ:tauT}) and (\ref{equ:TX}) are in agreement with the computations performed in similar models \cite{Carena:2006bn, Barbieri:2007bh}. The equivalent of equ.~(\ref{equ:tauX}) is not computed in those papers since the negative sign of the contribution~(\ref{equ:TX}) to $T$ is supposed to rule out the large-$m_{\tilde T}$ regime completely. A computation including this contribution has been recently performed by P.~Lodone \cite{Lodone:2008yy}, but our result do not agree with them. Notice however that in that very paper the realisation of the model is different from ours, in particular the global $SO(5)$ symmetry of the composite sector is not preserved.

The analytical results (\ref{equ:TT}) to (\ref{equ:tauQX}) show basically that the singlet $\tilde T$ gives positive contributions to both $\Delta\hat T$ and $\Delta \tau$, whereas the doublets $Q$ and $X$ give negative ones. Hence one should expect $\tilde T$ to be lighter than the doublets mass scale $M_0$ in order to obtain the desired positive contribution $\Delta\hat T$. However, the similitude of equations (\ref{equ:TT}) and (\ref{equ:tauT}) shows that a large $\Delta\hat T$ comes together with a large $\Delta \tau$, but the latter is strongly restricted by the experimental data. Therefore the allowed region of the parameter space may well be very thin.

\begin{figure*}[t]
	\centering
	\parbox{.48\linewidth}{\includegraphics[width=\linewidth]{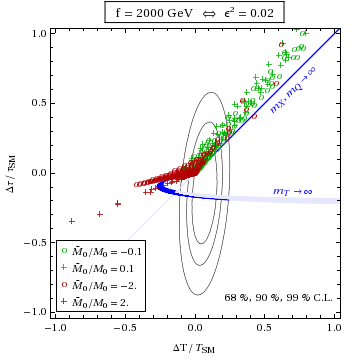}}
	\parbox{.48\linewidth}{\includegraphics[width=\linewidth]{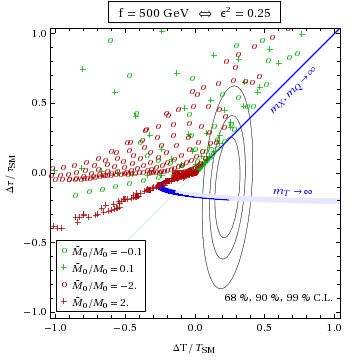}}
	\caption{Exact numerical computation of $\Delta T/T_\text{SM}^\text{top}$ and $\Delta \tau/\tau_\text{SM}^\text{top}$ for $f$ fixed to 2~TeV (left) and 500~GeV (right), and for four typical values of the ratio $\tilde M_0 / M_0$. The blue regions correspond to the analytical limits of equ.~(\ref{equ:TT}-\ref{equ:tauQX}), in light blue where the corresponding composite states are below 500~GeV (mostly irrelevant). The electroweak precision constraints are given by the ellipses.}
	\label{fig:dTdtau}
\end{figure*}

Indeed, performing an exact numerical computation, one can see that there is no possibility to obtain a significant positive contribution to $T$ and simultaneously to protect $\tau$ from growing too large. This statement is illustrated on fig.~\ref{fig:dTdtau}. However, it is still possible to get in agreement with the experimental precision measurements by having a small positive contribution to both $\Delta T/T^\text{SM}_\text{top}$ and $\Delta \tau/\tau^\text{SM}_\text{top}$. This is because the bounds on $\epsilon_b$ are weaker than those on $\epsilon_{1,2,3}$, since the correlation between the $\epsilon$'s given in equ.~(\ref{equ:epsilons}) is smaller for $\epsilon_b$. For example, with $f=500$~GeV or equivalently $\epsilon^2=0.25$, the model is excluded at about $4 \sigma$ if there are no contributions to $T$ and $\tau$ from new physics; however, a positive correction to both of them can drive the $\epsilon$'s into the $1\sigma$ confidence interval, as illustrated on the right-hand side of fig.~\ref{fig:dTdtau}. This precise computation of the electroweak precision observables was not performed in ref.~\cite{Barbieri:2007bh}, and thus their conclusion was that positive contribution to both $\Delta T$ and $\Delta \tau$ have to be excluded, whereas we have shown quantitatively that it can improve largely the agreement with experimental data. Notice that C.~Anastasiou et~al.~obtain similar numerical results 
\cite{Anastasiou:2009rv}.

\begin{figure*}[t]
	\centering
	\parbox{.48\linewidth}{\includegraphics[width=\linewidth]{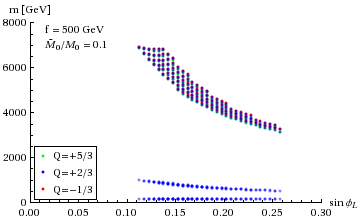}}
	\parbox{.48\linewidth}{\includegraphics[width=\linewidth]{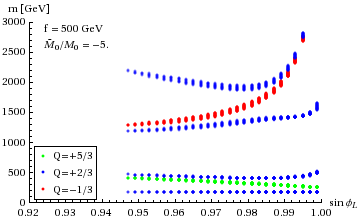}}
	\caption{Alloed mass of the fermions as a function of $\sin\phi_L$, with $f$ and the ratio $\tilde M_0/M_0$ fixed and $\phi_R$ varying freely. The green dots correspond to the charge $\frac{5}{3}$ quark $X^{5/3}$, the blue ones to the charge $\frac{2}{3}$ quarks $t$, $T$, $X^{2/3}$, $\tilde T$, and the red ones to the charge $-\frac{1}{3}$ quark $B$.}
	\label{fig:masses}
\end{figure*}

There are two ways of achieving this agreement, very different one from another. The first is simply to have a singlet $\tilde T$ much lighter than the rest of the composite sector: this occurs in the case of the cancellation $y^* f\sim-M_0$, which yields $\tilde M_0 \ll M_0$. Notice that in this case $M_0$ cannot be arbitrarily large since one should require $y^* < 4 \pi$ in order to be able to perform perturbative computations. On the other hand, the singlet should not be too light, due to the lack of direct experimental observation. For example, with $f=500$~GeV and $\tilde M_0/M_0=1/10$, a numerical computation shows that the singlet mass can run roughly between 500~GeV and 1~TeV, whereas the rest of the spectrum is situated in a mass range from 3 to 7~TeV (see fig.~\ref{fig:masses} left). The cancellation $y^* f\sim-M_0$ might be seen as a form of fine-tuning. However, the parameter $M_0$ is completely free in our model, and it is not unnatural to expect it to be of the same order of magnitude as the symmetry breaking scale $f$. Then since $y^*$ should be of order one, the cancellation can be achieved in a very natural way.

The alternative to such a large hierarchy between the singlet $\tilde T$ and the rest of the composite fermions is to have a very composite top quark, i.e.~to fix the mixing angle $\phi_L$ defined in equ.~(\ref{equ:phiL}) so that $\sin\phi_L \sim 1$. In this case the electroweak precision observables are in agreement with the experimental data even for a large $\tilde M_0/M_0$ ratio. The explanation for this fact is that the top Yukawa, given by equ.~(\ref{equ:topmass}) to be $\lambda_t = \sin\phi_L \sin\phi_R~y^*/\sqrt{2} + O(\epsilon^2)$, is fixed to the SM value (roughly one), and therefore $\sin\phi_R$ has to be smaller than one to compensate $\sin\phi_L \sim 1$ and $y^* \sim 4\pi$; but from equ.~(\ref{equ:TT}-\ref{equ:tauT}) one can see that a small $\sin\phi_R$ makes the contribution from the singlet dominant even if the doublet is light as well. An example of the corresponding physical masses is shown on fig.~\ref{fig:masses} right for $\tilde M_0/M_0=-5$: the doublet $X$ is very light, below 500~GeV, and the doublet $Q$ and the singlet $\tilde T$ have a heavier mass, between 1 and 3~TeV.

\begin{figure*}[t]
	\centering
	\parbox{.50\linewidth}{\includegraphics[width=\linewidth]{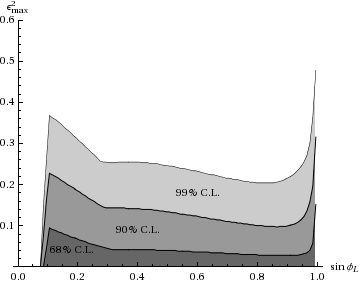}}
	\parbox{.46\linewidth}{\includegraphics[width=\linewidth]{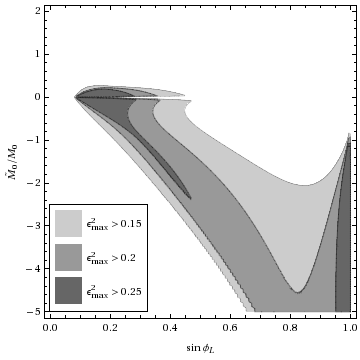}}
	\caption{Left: the maximal allowed value of $\epsilon^2$ as a function of $\sin \phi_L$, with $\sin \phi_R$ and $\tilde M_0/M_0$ varying freely. Right: regions of the plane $\sin \phi_L - \tilde M_0/M_0$ where $\epsilon^2$ can be larger than 0.15, 0.2 and 0.25 at 99\%~C.L.}
	\label{fig:epsmax}
\end{figure*}
The existence of those two regimes is illustrated by fig.~\ref{fig:epsmax} left, where we see that a large $\epsilon^2$ is allowed mainly in two regions: one for $\sin\phi_L \sim 0.1$, and one for $\sin\phi_L \sim 1$. It shows also that with an appropriate choice of the parameters, one can have $\epsilon^2$ to be as large as 0.35 (which corresponds to $f \sim 400$~GeV) and still agree with the electroweak precision constraints at 99\%~C.L. To have a better agreement with the experiment, however, the new physics should appear at a higher scale. Fig.~\ref{fig:epsmax} right shows the correlation between $\sin\phi_L$ and the ratio $\tilde M_0/M_0$. For 500~GeV~$\lesssim f \lesssim 1$~TeV, roughly any physical spectrum can be obtained since the ratio $\tilde M_0/M_0$ is not constrained anymore, but still a heavy singlet requires a large $\sin\phi_L$. For $f \gtrsim 1$~TeV, there is no real need for a positive contribution to $\Delta\hat T$ and we are back to the trivial case where all the new fermions are so heavy that they do not affect the EWPT at all.

Notice however that the large-$\sin\phi_L$ regime is problematic for flavour physics. For example, the mass difference $\Delta m_B$ of the neutral $B$ mesons is affected by the left-handed top compositeness. Requiring $\Delta m_B$ to be within the 20\% interval given by the experiment, one has roughly \cite{Giudice:2007fh}:
\begin{equation}
	\epsilon^2 \left( \sin\phi_L \right)^4 \lesssim 2 \cdot 10^{-3}.
\end{equation}
From this constraint, $\sin \phi_L$ can not be larger than 0.3 for $f=500$~GeV. $\sin \phi_L$ may actually affect the flavour physics in general, and therefore one should choose it as small as possible. The regime where the top is fully composite is therefore not realistic and should be excluded. Another strong constraint is coming from the $B \bar B$ mixing, since the corrections to the $\Delta B = 2$ operators is directly related to the correction to $\hat T$, as mentionned by Barbieri et~al. \cite{Barbieri:2007bh} In our case however the size of the positive contribution to $\hat T$ is still reasonable, and does not give rise to a significant constraint.

Since $\epsilon^2 = v^2/f^2$ is a measure of the fine-tuning of the model, it should not be much smaller than 10\%, i.e.~$f$ should not be significantly larger than 800~GeV. Since the top mass is directly proportional to $\epsilon$, $\sin \phi_L$ and $\sin \phi_R$ from equation~(\ref{equ:topmass}), the scale of the new physics becomes very large if all those parameters are small. Therefore one expects $\sin \phi_R$ to be of order one, which yields a fully composite right-handed top quark.

The characteristic signature of this class of composite models at LHC may come from the charge $5/3$ fermion, by single or pair production depending on its mass \cite{Contino:2008hi}. In any case, the discovery is made possible only if the charge $5/3$ quark has a mass roughly below 2~TeV.

The future improvements to the precision electroweak observables might change the results depicted in this paper. In particular if the present values are confirmed and the precision is improved, the regions of the parameter space away by 2 or 3 sigma from the preferred value might be excluded. However, it is important to note that the present data are nearly one sigma away from the Standard Model prediction, and therefore a composite Higgs model with $\epsilon^2 \lesssim 0.1$ is just as good as the Standard Model according to the EWPT.


\section{Conclusions}

We have computed quantitatively the corrections at one loop to the Peskin-Takeuchi precision parameters $S$ and $T$, and to the vertex of $Z \rightarrow b_L \bar b_L$ in the framework of a simple 4D two-site model \cite{Contino:2006nn} based on a 5D composite Higgs models \cite{Contino:2006qr,Carena:2006bn}. This class of models has to face important issues concerning the EPWT: the reduction of the coupling of the Higgs to the SM gauge fields leads to infrared corrections which affect the Peskin-Takeuchi $S$ and $T$ parameters, or similarly $\epsilon_1$ and $\epsilon_3$. In addition, a large ultraviolet correction due to the presence of a composite vector resonance is pulling this model outside the $1\sigma$ confidence level. Nevertheless, we have shown that in our particular model the constraints from the EWPT can be obtained within $2\sigma$ and $3\sigma$~C.L.~without a large amount of fine-tuning. More precisely, $\epsilon^2$ could be as large as 0.5 provided that $\sin \phi_L \sim 1$, i.e.~the left-handed top is fully composite. In this case, the new composite fermions would be at a mass scale accessible to the LHC. In particular the heavy charge $\frac{5}{3}$ quark should be produced significantly and give a characteristic signature of the model. But the full compositeness of the top quark is problematic for the flavour physics, so a full composite left-handed top is not realistic. The alternative is a regime where $\sin \phi_L \sim 0.1$, which requires a cancellation between the bare mass of the $SO(5)$ fermion $M_0$ and the Yukawa-like coupling $y^* f$. The latter option forces the composite doublets to be heavier than 3~TeV, but it yields a light singlet whose mass is below 1~TeV and can be nearly as light as the SM top quark. These requirements are relaxed when $\epsilon^2$ gets smaller. In summary, the EWPT do not rule out this particular model, but give strong bounds on the parameter space and on the corresponding physical spectrum of states.


\begin{acknowledgments}
	I am very thankful to Roberto Contino and Riccardo Rattazzi for their precious help on this work, as well as Jose Santiago, Elisabetta Furlan and Babis Anastasiou for useful discussions. This work was partially supported by the Swiss National Science Foundation.
\end{acknowledgments}




\end{document}